\documentclass[10pt,letterpaper]{article}
\usepackage[top=0.85in,left=2.75in,footskip=0.75in]{geometry}

\usepackage{changepage}

\usepackage[utf8]{inputenc}

\usepackage{textcomp,marvosym}

\usepackage{amsmath,amssymb}

\usepackage{cite}

\usepackage{nameref,hyperref}

\usepackage{microtype}
\DisableLigatures[f]{encoding = *, family = * }

\usepackage{rotating}

\usepackage{multirow,booktabs}
\usepackage{amsmath,amssymb,latexsym,MnSymbol}
\usepackage{hyperref}
\usepackage[table]{xcolor}

\raggedright
\usepackage[aboveskip=1pt,labelfont=bf,labelsep=period,justification=raggedright,singlelinecheck=off]{caption}

\makeatletter
\renewcommand{\@biblabel}[1]{\quad#1.}
\makeatother

\date{}

\usepackage{lastpage,fancyhdr,graphicx}
\usepackage{epstopdf}
\fancyheadoffset[L]{2.25in}
\fancyfootoffset[L]{2.25in}
\begin{document}
\vspace*{0.35in}

\begin{flushleft}
{\Large
\textbf\newline{Unveiling Relationships Between Crime and Property in England and Wales Via Density Scale-Adjusted Metrics and Network Tools}
}
\newline
\\
Haroldo V. Ribeiro\textsuperscript{1,*},
Quentin S. Hanley\textsuperscript{2},
Dan Lewis\textsuperscript{3},
\\
\bigskip
\bf{1} Departamento de F\'isica, Universidade Estadual de Maring\'a, Maring\'a, PR 87020-900, Brazil
\\
\bf{2} School of Science and Technology, Nottingham Trent University, Clifton Lane, Nottingham NG11 8NS, United Kingdom
\\
\bf{3} UkCrimeStats, Economic Policy Centre, London, SE1 3GA
\\
\bigskip

%
%





*\url{hvr@dfi.uem.br}

\end{flushleft}
\section*{Abstract}
Scale-adjusted metrics (SAMs) are a significant achievement of the urban scaling hypothesis. SAMs remove the inherent biases of \textit{per capita} measures computed in the absence of isometric allometries. However, this approach is limited to urban areas, while a large portion of the world's population still lives outside cities and rural areas dominate land use worldwide. Here, we extend the concept of SAMs to population density scale-adjusted metrics (DSAMs) to reveal relationships among different types of crime and property metrics. Our approach allows all human environments to be considered, avoids problems in the definition of urban areas, and accounts for the heterogeneity of population distributions within urban regions. By combining DSAMs, cross-correlation, and complex network analysis, we find that crime and property types have intricate and hierarchically organized relationships leading to some striking conclusions. Drugs and burglary had uncorrelated DSAMs and, to the extent property transaction values are indicators of affluence, twelve out of fourteen crime metrics showed no evidence of specifically targeting affluence. Burglary and robbery were the most connected in our network analysis and the modular structures suggest an alternative to ``zero-tolerance'' policies by unveiling the crime and/or property types most likely to affect each other.


\section*{Introduction}
Crime is a long-standing problem for society and its understanding has challenged scientists from a wide range of disciplines. From a sociological perspective, crime is treated as a deviant behavior of individuals and the goal of sociologists is often to find the conditions that lead to or favor criminal behavior. There is a vast literature on the sociology of crime seeking to find such conditions. An example is the ``broken windows theory''~\cite{Wilson1982} that correlates the incidence of crime with the existence of degraded urban environments. Despite the popularity and empirical support for this theory, there is a consensus that other factors than environment disorder are likely to affect or even have a greater influence on the incidence of crime. Situational action theory ~\cite{Wikstrom,Wikstrom2} seeks to understand how an individual's life history and social conditions interact with settings encouraging crime. More recently, crime has been considered as a complex system~\cite{Perc} where nonlinearities and self-organized principles create complex patterns that are difficult to understand and even harder to predict and control. This new perspective for studying crime and other social systems has been fostered by the availability of an unprecedented amount of data, making it possible to ask empirical questions that would have been considered unanswerable a few decades ago.

In the context of city-related metrics, researchers have recently promoted and made remarkable progress towards establishing the \emph{urban scaling hypothesis}~\cite{Pumain,Bettencourt,Arbesman,Bettencourt2,Bettencourt3,Mantovani,Gomez-Lievano,Bettencourt4,Lobo,Bettencourt5,Alves,Mantovani2,Alves2,OliveiraCO2,Alves3,Ignazzi,Louf2,Melo,RybskiCO2,Hanley,Rocha,Alves4,Hanley2,Pan,BettencourtN1,vanRaan,Youn,Fluschnik,Schlapfer,CaminhaHuman2017}. This theory states that cities are self-similar regarding their size as measured by population, meaning that several urban metrics (such as unemployment or a particular crime type) are expected to have a deterministic component that depends on the population of the city. The resulting scaling laws arise from only a few general assumptions about the properties of cities and should be universal across urban systems~\cite{Bettencourt5}. A consequence of these scaling laws is that \textit{per capita} measures are not appropriate for comparing urban units of different sizes and can exhibit biases favoring large or small cities depending on whether the relationship with the population is super or sublinear. In order to remove this bias, Bettencourt~\textit{et al.}~\cite{Bettencourt3,Lobo} proposed the use of a scale-adjusted metric (SAM) for removing the deterministic component associated with the population of an urban area. The SAMs are simply defined as the residuals of the fit to a scaling relationship between indicator and population. Despite their simplicity, SAMs can capture the exceptionality of a city regardless of its size and have proved useful for unveiling relationships that are not observed in \textit{per capita} measures~\cite{Lobo,Gomez-Lievano,Alves2,Alves4}.

The urban scaling hypothesis is supported by a wealth of empirical evidence using a wide range of urban indicators from many countries. However, the hypothesis has also been criticized~\cite{Masucci,Arcaute,Cottineau,Leitao} and one main criticism relates to the definition of the ``urban unity'' or city. Arcaute~\textit{et al.}~\cite{Arcaute} and Cottineau~\textit{et al.}~\cite{Cottineau} have shown that definitions of cities based on population density and commuter flows may lead to different observed scaling exponents. This challenges the idea that population size alone is responsible for the deterministic component of urban metrics and opens the possibility for other approaches.  In a recent article~\cite{Hanley2}, we argued that the relationship between an indicator density (\textit{e.g.} crime per hectare) and population density can provide a far superior framework when compared with traditional population scaling. In particular, this density-based approach is capable of continuously analyzing all human environments, from the most rural to heavily urban systems and identified that some metrics display scaling transitions at high population density, which can enhance, inhibit or even collapse the scaling exponents.

Here we further explore this density-based framework together with the scale-adjusted metrics approach to unveil relationships among different crime types and property values. Our approach extends the ideas of Bettencourt~\textit{et al.}~\cite{Bettencourt3,Lobo} by defining a density scale-adjusted metric (DSAM). In addition to removing the deterministic component, DSAMs enable the investigation of crime incidence and its relationships with property transaction values over the full range of human environments. Furthermore, by combining DSAMs, cross-correlation analysis, and complex network tools, we find that crime types have intricate and hierarchically organized relationships among themselves as well as with property values. Our approach reveals that these relationships are characterized by modular and sub-modular structures in which some crime types and/or property types are more likely to affect each other. 

\section*{Methods}

\subsection*{Data Sets}
The data set used in the present study is the same we have employed in Ref.~\cite{Hanley2}, where it is described in detail and made freely available (it has been also provided with this paper as \hyperref[S1_Dataset]{S1 Dataset}). Briefly, the data set consists of police-reported crimes, property transaction values, population size, and area for all 573 Parliamentary Constituencies in England and Wales. These data were collected on the UKCrimeStats (\url{http://www.ukcrimestats.com/}) data platform from different sources and subsequently reported as a snapshot since the data is regularly updated. Reported crimes are broken into 14 types while property data are categorized by 8 types (\autoref{tab:1}). 

\begin{table}[!ht]
\caption{\textbf{Constituency data analyzed in this study.}}
\begin{tabular}{c|l}
\hline
& Constituency metrics, $Y$ \\
\hline
\multirow{14}{*}{\rotatebox[origin=c]{90}{\parbox[c]{5cm}{\centering Crime Types}}} & Anti-Social Behavior (ASB) \cellcolor{gray!15} \\ 
& Bike Theft \\
& Burglary \cellcolor{gray!15} \\ 
& Criminal Damage and Arson (CD and A) \\
& Drugs \cellcolor{gray!15} \\
& Order \\
& Other Crime \cellcolor{gray!15} \\
& Other Theft \\
& Robbery \cellcolor{gray!15} \\
& Shoplifting \\
& Theft from the Person \cellcolor{gray!15} \\
& Vehicle Crime\\
& Violence \cellcolor{gray!15} \\
& Weapons\\
\hline
\multirow{8}{*}{\rotatebox[origin=c]{90}{\parbox[c]{3cm}{\centering Property Types}}} & Detached \cellcolor{gray!15} \\
& Flats \\
& Freehold \cellcolor{gray!15} \\
& Leasehold \\
& New \cellcolor{gray!15} \\
& Old \\
& Semi-detached \cellcolor{gray!15} \\
& Terraced \\
\hline
\multirow{2}{*}{\rotatebox[origin=c]{90}{\parbox[c]{1cm}{\centering }}} & Constituency population, $N$ \cellcolor{gray!15} \\
& Constituency area, $A$ \\
\hline
\end{tabular}
\label{tab:1}
\end{table}

\subsection*{Density Scaling Laws and Scale-Adjusted Metrics}
We start by revisiting the characterization of the density scaling laws previously described in Ref.~\cite{Hanley2}. The usual approach for studying urban scaling is by investigating the relationship between a given urban indicator $Y$ and population $N$ in a system composed of several ``urban units'' (such as municipalities). This relationship is often well described by a power-law relationship defined as
\begin{equation}\label{eq:allometry}
Y = Y_0 N^\beta~~\text{or its linearized version}~~\log Y = \log Y_0 + \beta \log N\,,
\end{equation}
where $Y_0$ is a constant and $\beta$ is the power-law or allometric exponent. In this context, urban indicators are categorized into three classes depending on whether the value of $\beta$ is equal (isometry), larger (superlinear allometry) or smaller (sublinear allometry) than 1. Metrics related to individual needs (\textit{e.g.} household energy and water consumption) usually have isometric relationships with population, while sublinear allometric relationships are observed for infrastructure metrics (\textit{e.g.} road surface and petrol stations) and superlinear allometric relationships appear for social, economic and health metrics (\textit{e.g.} crime, unemployment, and AIDS cases)~\cite{Bettencourt}. Thus, urban indicators have (in general) a nonlinear deterministic component associated with population. For a given city, this means that the value of a particular urban metric is expected to depend on the city's population in a nonlinear deterministic fashion.

A direct consequence of these nonlinearities is that \textit{per capita} measures are efficient in correctly removing the effect of population size in an urban metric only if the metric has an isometric relationship with the population. Otherwise, \textit{per capita} measures will be biased towards large populations (for superlinear allometries) or small populations (for sublinear allometries)~\cite{Alves4}. Consequently Bettencourt \textit{et al.}~\cite{Bettencourt3} defined the so-called \textit{scale-adjusted metric} (SAM). This metric consists of calculating the logarithmic difference between the actual value of an urban indicator and the value expected from the allometric relationship with population (Eq.~\ref{eq:allometry}); mathematically, we have (for the $i$-th city)
\begin{equation}\label{eq:SAM}
Z_i = \log Y_i - [\log Y_0 + \beta \log N_i]\,.
\end{equation}
It is worth noting that the scale-adjusted metric, $Z_i$, is the residual following the adjustment of an observation for the power-law defined by Eq.~\ref{eq:allometry}. The values of $Z_i$ capture the ``exceptionality'' of individual cities regarding a particular metric such that a positive/negative SAM indicates the metric is above/below the expectation for a city of that population. 

This approach has been successfully employed in economic and social contexts~\cite{Podobnik,Lobo,Alves2,Alves4} revealing relationships among metrics in urban systems which cannot be properly identified only by \textit{per capita} measures. In spite of its success, SAMs naturally share the same limitations of urban scaling. As previously mentioned, the allometric exponent depends on the definition of the ``urban unit'', and the urban scaling hypothesis is limited to urban areas by construction. On the one hand, the proportion of the world's population living in urban areas has been systematically increasing over the past decades and currently is around 54\%~\cite{UN}. On the other hand, the urbanization process is not uniform across all countries: there are countries where almost all the population is urban (such as Belgium and Uruguay where the proportion of urban population is larger than 95\%) while others are predominantly  rural (such as India with 33\% of urban population and Trinidad and Tobago with only 9\%)~\cite{WorldBank}. Furthermore, in countries where most of the population is urban, rural areas may represent the vast majority of the countries' land. The United Kingdom is one such country with a population that is  83\% urban but rural areas cover 85\% of the land~\cite{GovtStatServ}. Thus, it is important to develop a framework capable of investigating the full range of human environments.

Previously, we proposed an approach for taking these problems into account~\cite{Hanley2}. Our idea was to analyze scaling relationships between an indicator density and population density over all 573 parliamentary constituencies of England and Wales, regions that range in population density from very rural ($0.22$~p/ha) to heavily urban ($550.3$~p/ha). In place of Eq.~\ref{eq:allometry}, we considered the following generalization (see also~\cite{hanley2016correction})
\begin{equation}\label{eq:allometry_den}
\log y =
\begin{cases}
\log y_0 + \beta_L \log d & \text{for}~~d<d^*\\
\log y_1 + \beta_H \log d & \text{for}~~d\geq d^*
\end{cases}\,,
\end{equation}
where $y = Y/A$ is the indicator density, $d= N/A$ is the population density, $d^*$ is a population density threshold at which the allometric exponent changes from $\beta_L$ to $\beta_H$, $y_0$ and $y_1$ are constants ($\log y_1 = \log y_0 + (\beta_L -\beta_H) \log d^*$ for holding the continuity of the indicator density $y$ as a function of the population density $d$). For crime and property metrics, we found this approach to be superior to traditional population scaling, significantly refining our understanding of scaling laws in human environments. Depending on the metric, we have found rural-to-urban transitions with thresholds in the range of $10$-$70$ people per hectare. These transitions were characterized by enhanced, inhibited, or collapsed scaling at high population densities. For some metrics, a single continuous relationship was observed from the most rural to heavily urban environments; that is, Eq.~\ref{eq:allometry_den} is reduced to a single power-law relationship
\begin{equation}\label{eq:allometry_den2}
\log y = \log y_0 + \beta \log d\,,
\end{equation}
where $\beta$ is the allometric exponent of the density scaling law.

Within this framework and by following the ideas of Bettencourt \textit{et al.}~\cite{Bettencourt3}, we define the \textit{density scale-adjusted metrics} (DSAMs) $z_i$ (for the $i$-th constituency) as a direct generalization of Eq.~\ref{eq:SAM} in the context of the density scaling laws, that is,
\begin{equation}\label{eq:DSAM}
z_i = \log y_i - f(d_i)\,,
\end{equation}
where 
\begin{equation}
f(d) = 
\begin{cases}
\log y_0 + \beta_L \log d & \text{for}~~d<d^*\\
\log y_1 + \beta_H \log d & \text{for}~~d\geq d^*\,
\end{cases}\,,
\end{equation}
when the metric has a rural-to-urban transition at the population density threshold $d^*$, and  
\begin{equation}
f(d) = \log y_0 + \beta \log d\,,
\end{equation}
when the metric behaves continuously with the population density. 

\begin{figure*}[!ht]
\begin{adjustwidth}{-2.25in}{0in}
\begin{center}
\includegraphics[scale=0.4]{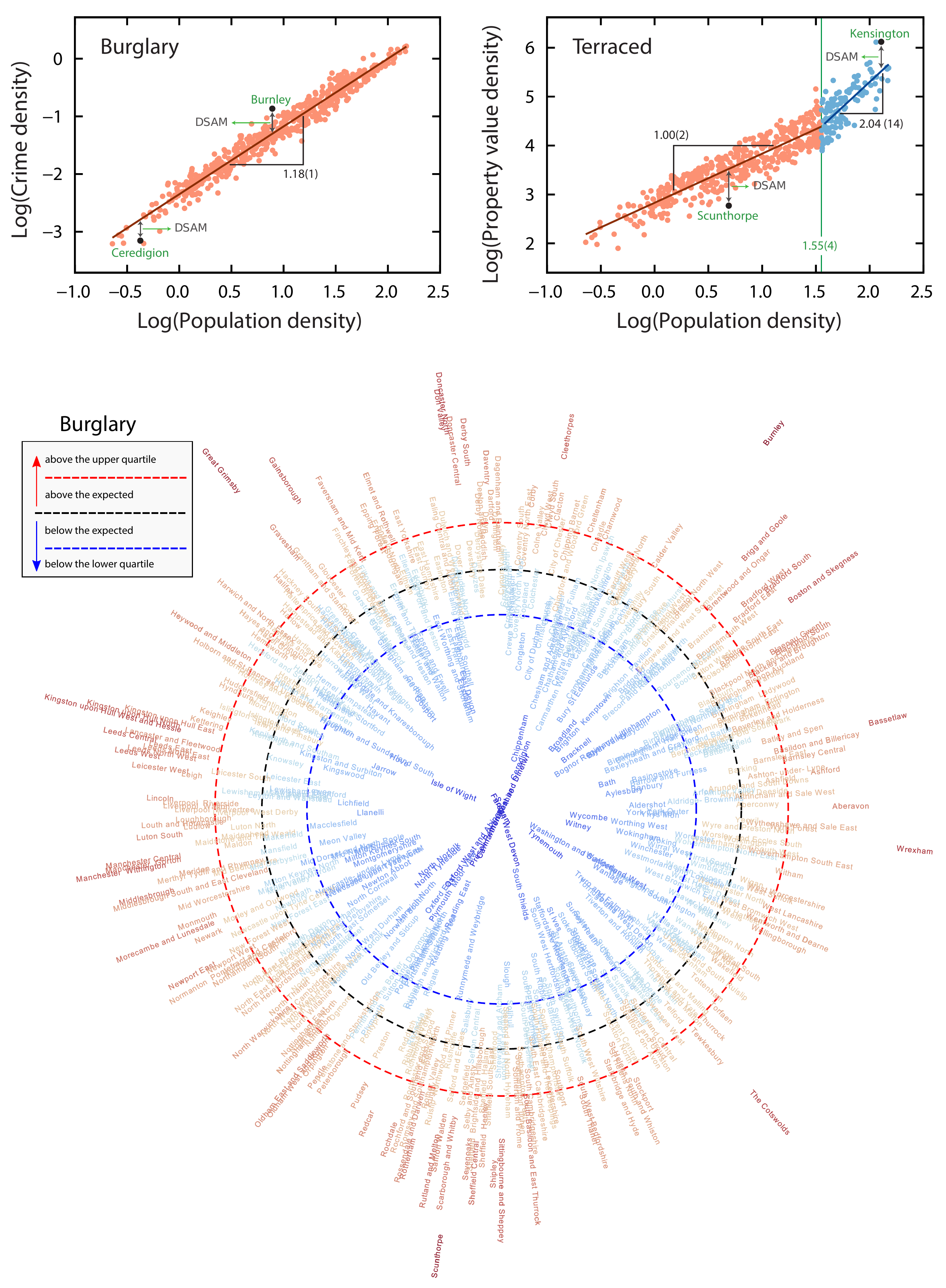}
\end{center}
\caption{
\textbf{Allometric laws between indicator density and population density, and the definition of density scale-adjusted metrics (DSAMs).} Upper panels illustrate how four DSAMs (Ceredigion, Burnley, Scunthorpe, and Kensington) are generated using the scaling laws from Ref.~\cite{Hanley2}. The relationship between indicator density and population density for a crime (burglary) and property (terraced) type on a log-log scale. For burglary, a simple allometric law with an exponent $\beta=1.18(1)$ fits this relationship; whereas for terraced, two power laws ($\beta_L=1.00(2)$ low population density and $\beta_H=2.04(14)$ for high) is a better description. The continuous black lines are the adjusted power laws and the green vertical line indicates the transition between low and high densities for terraced properties ($\log d^* = 1.55(4)$). The density scale-adjusted metric (DSAM) is defined as the difference (in logarithmic scale) between the value of an indicator density and the value that is expected based on population density via the power-law fit (single or double). Arrows in these plots illustrate the DSAM for four Parliamentary Constituencies: Ceredigion, Burnley, Scunthorpe, and Kensington. The bottom panel shows a visualization of the DSAMs for burglary for all Parliamentary Constituencies. The constituency names are disposed in alphabetic order (counterclockwise direction) in a radial plot, in which the distances to central point are proportional to the Constituencies' DSAM. Constituencies within the black dashed circle have burglary density lower than the expected by their population density (negative DSAM, azure shades); whereas those outside this circle have burglary density higher than the expected (positive DSAM, red shades). The blue and the red dashed circles indicate the lower and upper quartile of the DSAM distribution (see~\hyperref[S1_file]{File S1} for all other indicators). 
}
\label{fig:1}
\end{adjustwidth}
\end{figure*}

All best fit parameters for $y_0$, $y_1$, $\beta_L$, $\beta_H$, $d^*$, and $\beta$ are available in Ref.~\cite{Hanley2} and reproduced in \autoref{tab:2}. Analogously to SAMs, the values of $z_i$ are simply the residuals surrounding the allometric relationships between an indicator density and population density, as illustrated in Figure~\ref{fig:1} (upper panels) for burglary and terraced housing using scaling laws from Ref.~\cite{Hanley2}. The DSAMs are continuous and for a particular indicator can be placed into quartiles as has been done for burglary (Figure~\ref{fig:1}, lower panel -- see also~\hyperref[S1_file]{File S1} for all indicators). This allows regions having exceptionally high (\textit{e.g.} Burnley) or low (\textit{e.g.} Ceredigion) DSAMs to be identified. It should be noted that as residuals from a scaling law, these are not absolute metrics of crime density or property values, but indicate how a particular constituency compares to predicted behavior based on population density. The absolute values for burglary in constituencies such as Burnley, the Cotswalds, Scunthorpe, and Greater Grimsby would not appear as exceptional in the absence of scale adjustment. However, these areas all exhibited exceptionally high burglary density relative to expected values.   

\begin{table}[!ht]
\begin{adjustwidth}{-2.25in}{0in}
\caption{\textbf{Best fitting parameters $\log y_0$, $\beta_L$, $\beta$, $\log y_1$, $\log d^*$, and $\beta_H$ of models described in Eqs.~\ref{eq:allometry_den} and~\ref{eq:allometry_den2} for each crime and property type.} Values reproduced from Ref.~\cite{Hanley2}.}
\centering
\begin{tabular}{lrrrrr}
\hline
{Crime Type} & \multicolumn{1}{c}{$\log(y_0)$} & \multicolumn{1}{c}{$\beta_{\text{L}}$ or $\beta$} & \multicolumn{1}{c}{$\log(y_1)$} & \multicolumn{1}{c}{$\log(d^*)$} & \multicolumn{1}{c}{$\beta_{\text{H}}$}\\
\hline
\rowcolor{gray!15} ASB & $-1.62\pm0.02$ & $1.13\pm0.02$ & $-1.30\pm0.13$ & $1.47\pm0.13$ & $0.91\pm0.08$ \\ 
Bike Theft & $-3.26\pm0.02$ & $1.27\pm0.02$ & $-4.62\pm0.77$ & $1.80\pm0.12$ & $2.03\pm0.43$\\
\rowcolor{gray!15} Burglary & $-2.35\pm0.01$ & $1.18\pm0.01$ & \multicolumn{1}{c}{-} & \multicolumn{1}{c}{-} & \multicolumn{1}{c}{-}\\
CD and A & $-2.21\pm0.01$ & $1.14\pm0.01$ & $-1.55\pm0.11$ & $1.52\pm0.05$ & $0.71\pm0.07$\\
\rowcolor{gray!15} Drugs & $-2.77\pm0.02$ & $1.08\pm0.03$ & $-3.13\pm0.08$ & $1.13\pm0.10$ & $1.40\pm0.05$\\
Order & $-2.91\pm0.02$ & $1.16\pm0.03$ & $-3.20\pm0.07$ & $1.06\pm0.12$ & $1.43\pm0.05$\\
\rowcolor{gray!15} Other Crime & $-3.29\pm0.01$ & $1.15\pm0.01$ & \multicolumn{1}{c}{-} & \multicolumn{1}{c}{-} & \multicolumn{1}{c}{-}\\
Other Theft & $-2.26\pm0.01$ & $1.11\pm0.01$ & $-2.57\pm0.08$ & $1.40\pm0.09$ & $1.33\pm0.05$\\
\rowcolor{gray!15} Robbery & $-3.98\pm0.02$ & $1.55\pm0.03$ & $-4.73\pm0.14$ & $1.32\pm0.08$ & $2.12\pm0.10$\\
Shoplifting & $-2.56\pm0.02$ & $1.26\pm0.02$ & $-1.61\pm0.16$ & $1.50\pm0.06$ & $0.63\pm0.10$\\
\rowcolor{gray!15} Theft from the Person & $-3.68\pm0.03$ & $1.36\pm0.03$ & $-4.84\pm0.18$ & $1.39\pm0.06$ & $2.20\pm0.12$\\
Vehicle Crime & $-2.54\pm0.01$ & $1.27\pm0.01$ & \multicolumn{1}{c}{-} & \multicolumn{1}{c}{-} & \multicolumn{1}{c}{-}\\
\rowcolor{gray!15} Violence & $-2.06\pm0.01$ & $1.12\pm0.02$ & $-2.28\pm0.06$ & $1.17\pm0.13$ & $1.30\pm0.04$\\
Weapons & $-3.78\pm0.02$ & $1.23\pm0.02$ & \multicolumn{1}{c}{-} & \multicolumn{1}{c}{-} & \multicolumn{1}{c}{-} \\
\hline
Property Type & & & & &\\
\hline
\rowcolor{gray!15} Detached & $3.30\pm0.03$ & $0.77\pm0.04$ & $4.47\pm0.14$ & $1.21\pm0.06$ & $-0.20\pm0.10$\\
Flats & $2.13\pm0.05$ & $1.13\pm0.05$ & $-1.65\pm0.48$ & $1.55\pm0.04$ & $3.57\pm0.30$\\
\rowcolor{gray!15} Freehold & $3.55\pm0.02$ & $0.83\pm0.02$ & $2.48\pm0.42$ & $1.70\pm0.10$ & $1.46\pm0.25$\\
Leasehold & $2.24\pm0.04$ & $1.26\pm0.04$ & $-1.83\pm0.69$ & $1.68\pm0.04$ & $3.68\pm0.40$\\
\rowcolor{gray!15} New & $2.30\pm0.03$ & $0.86\pm0.03$ & $-1.88\pm1.06$ & $1.80\pm0.05$ & $3.19\pm0.58$\\
Old & $3.55\pm0.02$ & $0.89\pm0.02$ & $0.92\pm0.42$ & $1.71\pm0.04$ & $2.43\pm0.24$\\
\rowcolor{gray!15} Semi Detached & $2.90\pm0.02$ & $1.05\pm0.03$ & $3.84\pm0.14$ & $1.41\pm0.06$ & $0.38\pm0.09$\\
Terraced & $2.83\pm0.02$ & $1.00\pm0.02$ & $1.23\pm0.22$ & $1.55\pm0.04$ & $2.04\pm0.14$\\
\hline
\end{tabular}
\label{tab:2}
\end{adjustwidth}
\end{table}

\section*{Results and Discussion}
\subsection*{Statistical Properties of DSAMs}
\begin{figure*}[!ht]
\begin{adjustwidth}{-2.25in}{0in}
\begin{center}
\includegraphics[scale=0.43]{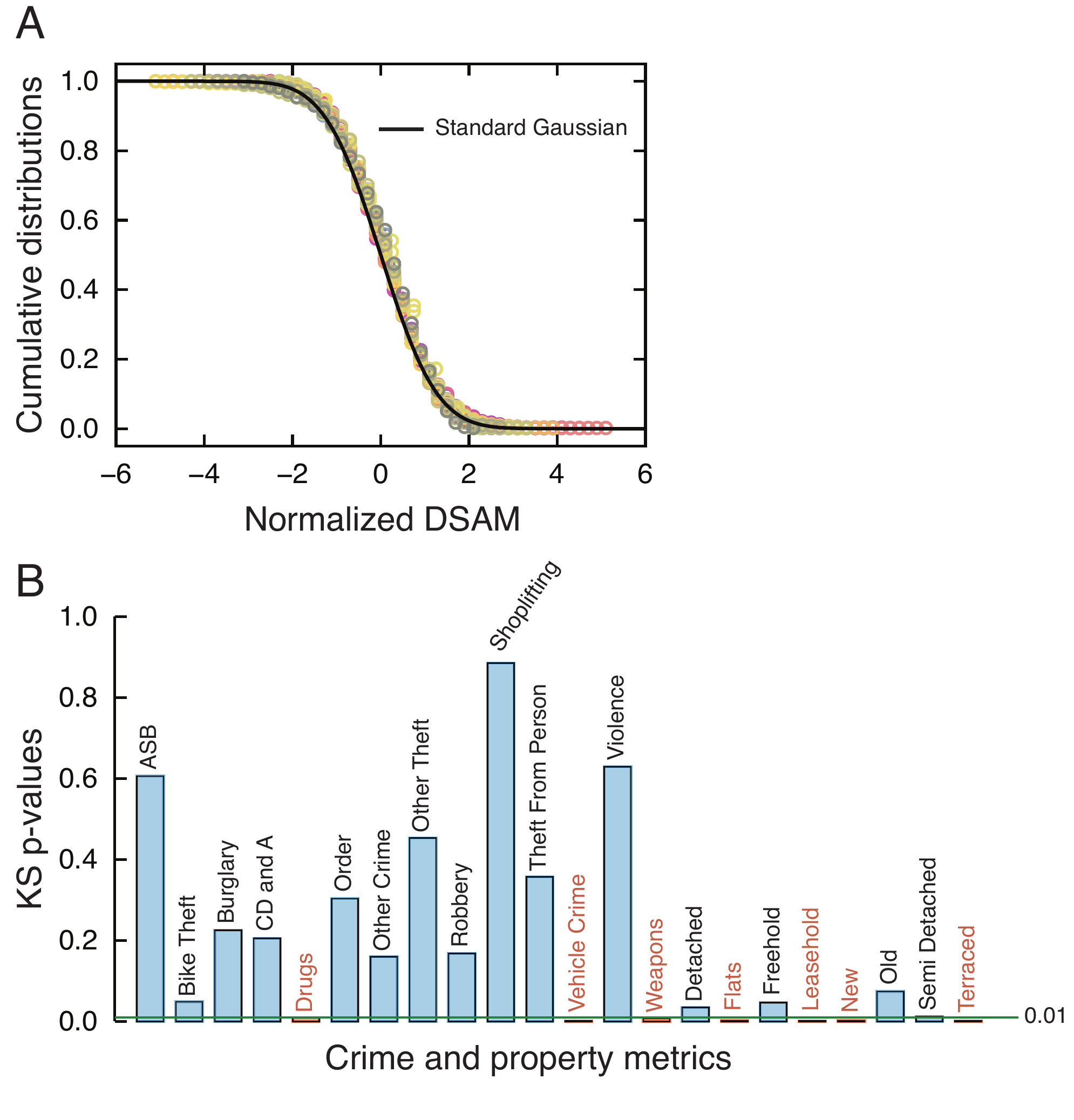}
\end{center}
\caption{
\textbf{Most DSAMs are normally distributed.} (a) Cumulative distribution of the normalized DSAM (that is, after subtracting its mean and dividing by its standard deviation) for each crime and property type (colored circles). We note that all distributions are very close to the standard Gaussian (zero mean and unitary variance) indicated by the continuous line. (b) The $p$-values of the Kolmogorov-Smirnov normality test for each crime and property type. The normality of DSAMs cannot be rejected at 99\% confidence level for 15 of the indicators (blue bars), whereas this hypothesis is rejected for drugs, vehicle crime, weapons, flats, leasehold, new and terraced (red bars, 7 indicators). See~\hyperref[S1_Fig]{S1 Fig} for individual distributions and for a comparison between single and double power-law model for calculating the DSAMs.}
\label{fig:2}
\end{adjustwidth}
\end{figure*}

Having defined DSAMs, it is useful to study their statistical properties. We note that DSAMs have been defined in a log-log domain and ask whether their values are distributed according to a Gaussian distribution, a feature that facilitates the use of linear regression in forecast analysis~\cite{Alves3,Alves4}. To do so, we set $w=15$ equally-spaced windows over the logarithm of the population density and within each one we calculate the average ($\mu_w$) and the standard deviation ($\sigma_w$) of the DSAMs. The normalized DSAMs are defined by subtracting (within each window) the mean $\mu_w$ from $z_i$ and dividing the result by the standard deviation $\sigma_w$. This gives a standardized score (or $z$-score) calculated within windows of population density. It worth noting that $\mu_w$ is very close to zero due to the DSAM definition and thus subtracting $\mu_w$ from $z_i$ or not is irrelevant for the results we present in this section. Also, our results are very robust for different number of windows $w$; in particular, our conclusions are not modified if $8\leq w\leq20$. Figure~\ref{fig:2}A shows the normalized DSAM cumulative distributions for all crime and property types in comparison with the standard Gaussian, where a good correspondence is observed. 

A more rigorous analysis of the normality of the DSAMs is obtained by applying the Kolmogorov-Smirnov (KS) test~\cite{corder2014nonparametric} to the values of $z_i$. Figure~\ref{fig:2}B shows the $p$-values of the test for each indicator, where we observe that normality cannot be rejected for the majority of the indicators (15 out of 22); however, the normality of four property types (flats, leasehold, new, and terraced properties) and three crime categories (drugs, vehicle crime, and weapons) could be rejected with 99\% confidence. This indicates that these property and crime types have a more complex dependence on population density (as also discussed in~\cite{Leitao} in the context of the population scaling). One possibility is that some of these indicators have additional unobserved population density thresholds which our sample size was insufficient to detect. We suspect this due to comparing the normality of standardized DSAMs from single power-law (Eq.~\ref{eq:allometry_den2}) and double power-law models (Eq.~\ref{eq:allometry_den}) and finding that the $p$-values of the KS tests always improved (when below confidence level -- see~\hyperref[S1_Fig]{S1 Fig}) when considering the double power-law model. In particular, when a double power-law model was applied, normality can no longer be rejected for the indicators criminal damage and arson (CD and A), detached, freehold, and semi detached (\hyperref[S1_Fig]{S1 Fig}). Despite this possibility, normality is observed for most indicators and in the exceptions the deviations are not large allowing us to consider, to a first approximation, DSAMs to be normally distributed. It is worth mentioning that the normality of $z_i$ implies that the fluctuations in crime and property densities are distributed according to a log-normal shape, as also observed in the usual population scaling~\cite{Alves3,Gomez-Lievano,Alves2,Alves4,Mantovani,Mantovani2}.

Another interesting question regarding DSAMs is whether their fluctuations increase with the population density. This question is related to Taylor's law~\cite{taylor,Eisler}, which establishes an empirical power-law relationship between the variance (or standard deviation) and the ensemble average over groups of similar size (this law can also be applied to time averages~\cite{Eisler}). In population scaling this parallel is more direct, since one can consider groups of similar size to be cities with similar population. Empirical results from population scaling have found no (strong) evidence that the fluctuations surrounding scaling laws increase with population size in a logarithmic space~\cite{Alves3,Gomez-Lievano,Alves2,Alves4,Mantovani,Mantovani2}. 

\begin{figure}[!ht]
\begin{adjustwidth}{-2.25in}{0in}
\begin{center}
\includegraphics[scale=0.13]{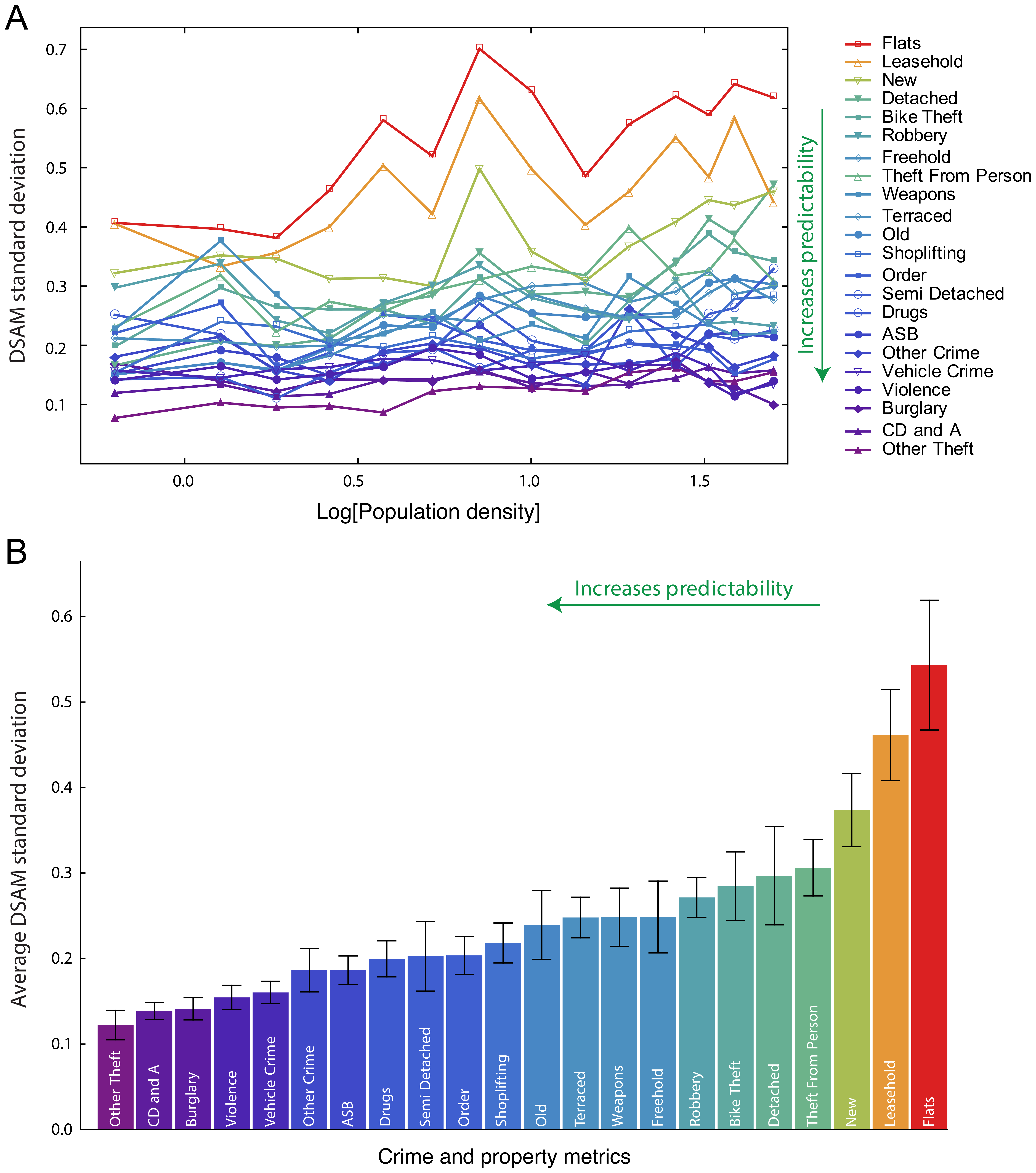}
\end{center}
\caption{
\textbf{Fluctuations of the DSAMs are independent of log population density.} (a) Standard deviation of DSAM versus population density (on a log scale) for all crime and property types. These values are evaluated in 15 equally-spaced windows over population density on a logarithmic scale. The behavior of these curves is well approximated by constant plateaus (see~\hyperref[S2_Fig]{S2 Fig} for individual plots). Linear regression found significant linear increasing trends for 9 indicators; however, the growth rates are very small in all cases (up to 0.005 units of DSAM per log[p/ha] -- see~\hyperref[S3_Fig]{S3 Fig}). (b) Average value of the DSAM standard deviation for each crime and property type in ascending order. Error bars represent 99\% bootstrap confidence intervals. As the standard deviation decreases, the accuracy increases in the prediction of the indicator based on the population density.}
\label{fig:3}
\end{adjustwidth}
\end{figure}

In our case, we consider that constituencies with similar population densities form groups for studying the dependence of the standard deviation of crime and property DSAMs on the population density (note: this corresponds to the relationship between log($\sigma_w$) and log($d$)). Figure~\ref{fig:3}A depicts this analysis for all indicators, where no clear dependence between the DSAM standard deviation and the logarithm of population density is observed (see also \hyperref[S2_Fig]{S2 Fig} for individual plots). By using linear regression~\cite{corder2014nonparametric}, we find no significant linear trend in 13 out of 22 indicators; furthermore, the growth rates of $\sigma_w$ for the indicators showing significant trends are very small (up to 0.005 units of DSAM per log[p/ha] -- \hyperref[S2_Fig]{S2 Fig} and \hyperref[S3_Fig]{S3 Fig}). 

Thus, our results are similar to those reported for population scaling, that is, there is limited evidence supporting the hypothesis that the fluctuations surrounding density scaling laws increase with the population density in the log-log space. As discussed in Leitão \textit{et al.}~\cite{Leitao}, this constant variance reinforces the conclusion that DSAMs are approximately normally distributed, subject to the caveat that where normality was rejected it is likely that a better model could be found. In Leitão \textit{et al.}, they proposed a probabilistic framework for hypothesis testing that explicitly considers the fluctuations for fitting allometric laws with population. Within this approach, they concluded that most of their models could be rejected in hypothesis testing, which also suggests that more complex patterns (such as the double power-law model employed here) are present in population scaling. Also, this nearly constant behavior allows us to characterize the deviations from the predictions of the density scaling laws by the average value of the DSAMs standard deviation over all population densities. The smaller the value of this quantity, the more predictable the metric is in terms of the population density. Figure~\ref{fig:3}B shows a bar plot of this quantity for all indicators in ascending order, where we observe that crime types usually have smaller fluctuations than property metrics.

\subsection*{DSAM Cross-Correlations}
Having all DSAMs calculated and because this approach efficiently removes the deterministic effect of population density, we can now ask about inter-relationships between the different crime and property categories. To do so, we first ask whether DSAMs are cross-correlated in a linear fashion. For SAMs, researchers have reported that in addition to removing the population bias, these metrics have linear correlations among each other~\cite{Bettencourt3,Podobnik,Lobo,Alves4}. In the DSAMs case, we tested this hypothesis by estimating the maximal information coefficient (MIC)~\cite{Reshef}. This non-parametric coefficient measures the association between two variables, even when they are correlated in nonlinear fashions. The MIC also coincides to the square of the Pearson linear correlation~\cite{Reshef} for linear relationships; thus, the difference between MIC and the square of the Pearson correlation is a natural measure of the degree of nonlinearity between two variables~\cite{Reshef}. Because of that, we estimate the MIC ($\mathcal{M}_{ij}$) and the Pearson correlation coefficient~\cite{corder2014nonparametric} ($\rho_{ij}$) for every possible pair of DSAM types ($i$ and $j$), and calculate the average value of the difference $\mathcal{M}_{ij} - \rho_{ij}^2$ over all unique DSAM type pairs ($i>j$). We further calculate this average when random shuffling of the DSAMs among the constituencies and for a set of uniform random variables with size equal to number constituencies. \hyperref[S4_Fig]{S4 Fig} shows that the average of the difference $\mathcal{M}_{ij} - \rho_{ij}^2$ for the original DSAM set is small ($0.09\pm0.06$) and not significantly different from the averages calculated from the shuffled DSAMs and random variables. We have also tested the linearity of the DSAMs relationships by comparing the AIC (Akaike information criterion~\cite{Burnham}) values of linear models adjusted to these relationships with those obtained from quadratic and cubic models. To do so, we bootstrap the AIC values among all possible pairs of DSAMs and test whether the difference is significant by using the two-sample bootstrap mean test. Results show that quadratic relationships are better descriptions (compared with linear) only in 8\% all pairwise relationships; similarly, cubic relationships are better models only in 10\% of cases. Therefore, in addition to removing the effect of population density, the DSAMs from each type of metric also show linearly correlation among each other.  

\begin{figure*}[!ht]
\begin{adjustwidth}{-2.25in}{0in}
\begin{center}
\includegraphics[scale=0.56]{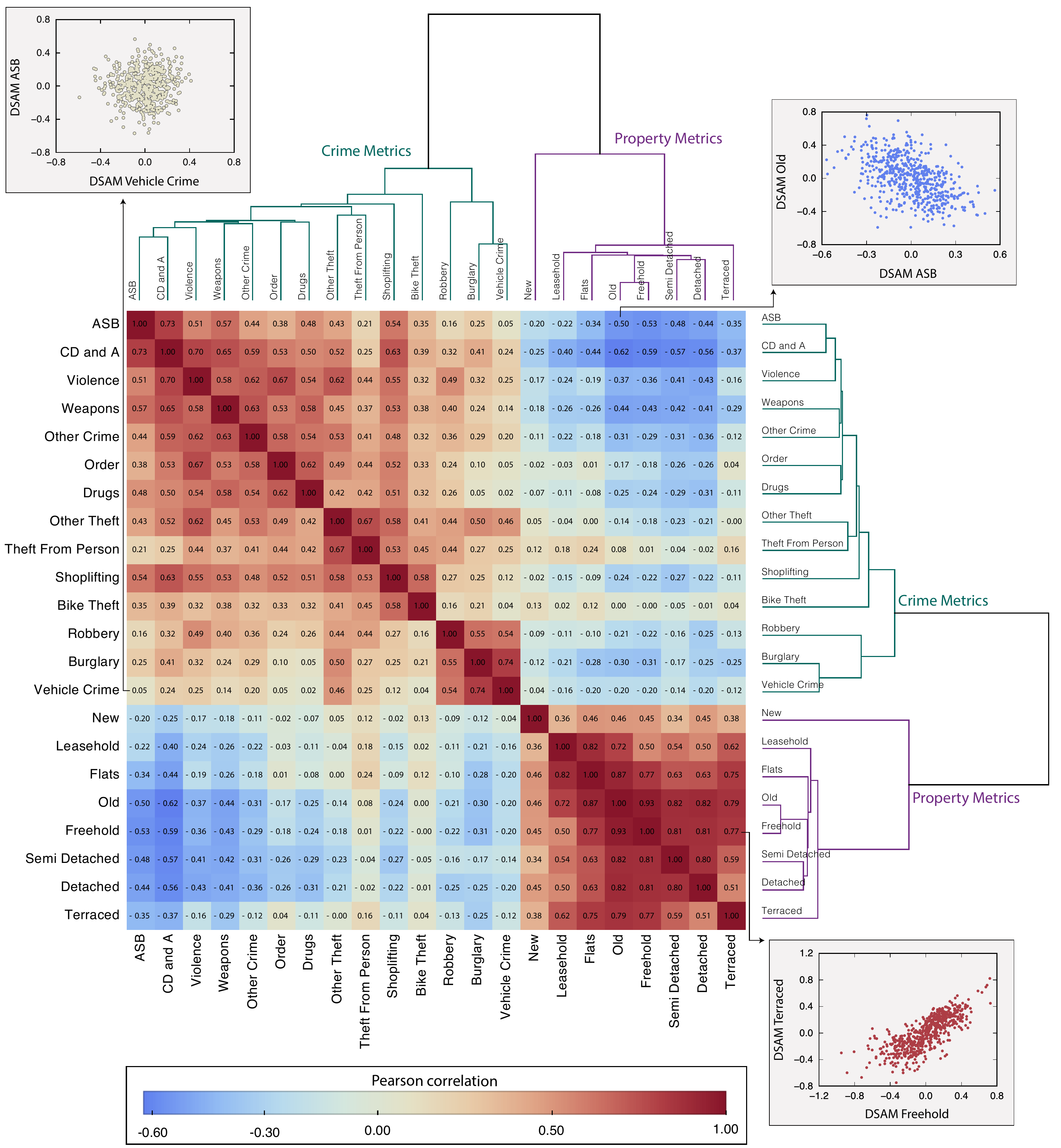}
\end{center}
\caption{
\textbf{Crime and property DSAMs are cross-correlated and form a hierarchical structure.} The matrix plot shows the value of the Pearson correlation coefficient ($\rho_{ij}$) evaluated for each combination of crime and property DSAM ($i$ and $j$). The number inside each cell is the coefficient value and the color code also refers to $\rho_{ij}$ (blue indicates negative correlation, while red is used for positive correlations; the darker the shade, the stronger the correlation). The insets indicated by arrows show examples of relationships among crime and property DSAMs. Upper and right-side panels are dendrograms constructed via the hierarchical clustering algorithm (based on the distance $d_{ij}=\sqrt{2(1-\rho_{ij})}$).
}
\label{fig:4}
\end{adjustwidth}
\end{figure*}

Figure~\ref{fig:4} shows the correlation matrix $\rho_{ij}$ for every possible pair of DSAM ($i$ and $j$). In order to better understand these inter-relationships, we define the ultrametric distance matrix $d_{ij}=\sqrt{2(1-\rho_{ij})}$ for applying the single-linkage clustering algorithm~\cite{mantegna1999introduction}, yielding the dendrograms shown in Figure~\ref{fig:4}. Several conclusions are clear from inspection of this figure: 
\begin{itemize}

\item For all property types there is a positive correlation in property transaction value DSAMs with those of all other property types. Most were very strong with many above $0.7$ with values reaching $0.93$ (old vs. freehold). Positive correlations indicate the tendency for high values of one property type to be associated with high values in all other property types.

\item All crime types are positively correlated with all other crime types with some strong correlations (\textit{e.g.} 0.73 for anti-social behavior vs. criminal damage and arson -- ASB vs. CD and A). In contrast to property types, the correlations among crime metrics were not as strong and some were very weak with insignificant correlation (\textit{e.g.} 0.02 for vehicle crime vs. drugs).

\item The only anti-correlations seen are between crime and property DSAMs. This gives rise to the blue regions in the upper right and lower left regions of Fig 4. Anti-correlations indicate the tendency for a positive property DSAM to be associated with a negative crime DSAM (\textit{e.g.} high property value DSAM is associated with low crime). The majority of crime vs. property DSAMs are anti-correlated which demonstrates a tendency for crime to be associated with depressed property transaction values. The three strongest predictors of depressed property value DSAMs were criminal damage and arson (CD and A), anti-social behavior (ASB), and weapons with old and freehold properties most affected. This does not prove crime as the causative agent, but does demonstrate the association over a wide range of indicators. 

\item Two crime types (theft from the person and bike theft) exhibited positive crime vs. property correlations. This is a good example to illustrate that one has to be careful when trying to associate causal relationships to these correlations. If taken literally, one could absurdly think that to improve property values, we must encourage bike theft and theft from the person. A more logical explanation is that these two crime types tend to rise in regions of relative affluence, assuming that property transaction value DSAMs are metrics of relative affluence. Again, this does not prove causation, however, it does make clear that it is only these 2 (out of 14) crime types which show any evidence of being attracted to or specifically targeting affluence.

\end{itemize}

The hierarchical clustering behavior reinforced many of these conclusions. We note the emergence of two main clusters setting apart crime and property metrics. In the property data, new property appears isolated from the remaining property types. This is a striking result because, with the exception of old property, every property category examined can include new properties as the classifications are not exclusive. Within the crime metrics, there is a sub-cluster consisting of robbery, burglary, and vehicle crime distinct from other crime types. The remaining crime types form a separate group with an important sub-cluster consisting of anti-social behavior (ASB), criminal damage and arson (CD and A), and violence. Interestingly, despite considerable discussion of drugs and burglary in the literature~\cite{Cromwell, Kuhns2017}, drugs and burglary crime reports are uncorrelated in our data. This discrepancy may be due to the design of many previous studies in which convicted offenders are surveyed. It is likely that drug use contributes to burglars being apprehended and convicted. Hence, the subset of all burglars composed of known offenders may not be representative of burglars in general. In our data, drugs crime reports are much more strongly associated with reports of order and weapons offenses.   

\subsection*{DSAM Networks}
Another approach for probing patterns in the complex inter-relationships among crime and property metrics is to create a complex network representation~\cite{newman2010networks,albert2002statistical}. The hierarchical classification was able to distinguish the difference between the crime and property metrics clearly and also identify subcategories. This representation works well for positive correlations, but failed to identify the important negative correlations between certain types of crime and property. In addition, the two dimensional grid structure limits the number of neighbors that can be placed adjacent to a particular category, and the dendrogram does not account for strength or significance of the correlations. Furthermore, complex networks (or spaces) already have shown very to be quiet useful to understand how several socioeconomic phenomena are related to each other~\cite{hidalgo2007product,brascoupe2010causes,neffke2011regions,caldarelli2012network,neffke2013skill,muneepeerakul2013urban,yildirim2014using}.

\begin{figure*}[!ht]
\begin{adjustwidth}{-2.25in}{0in}
\begin{center}
\includegraphics[scale=0.22]{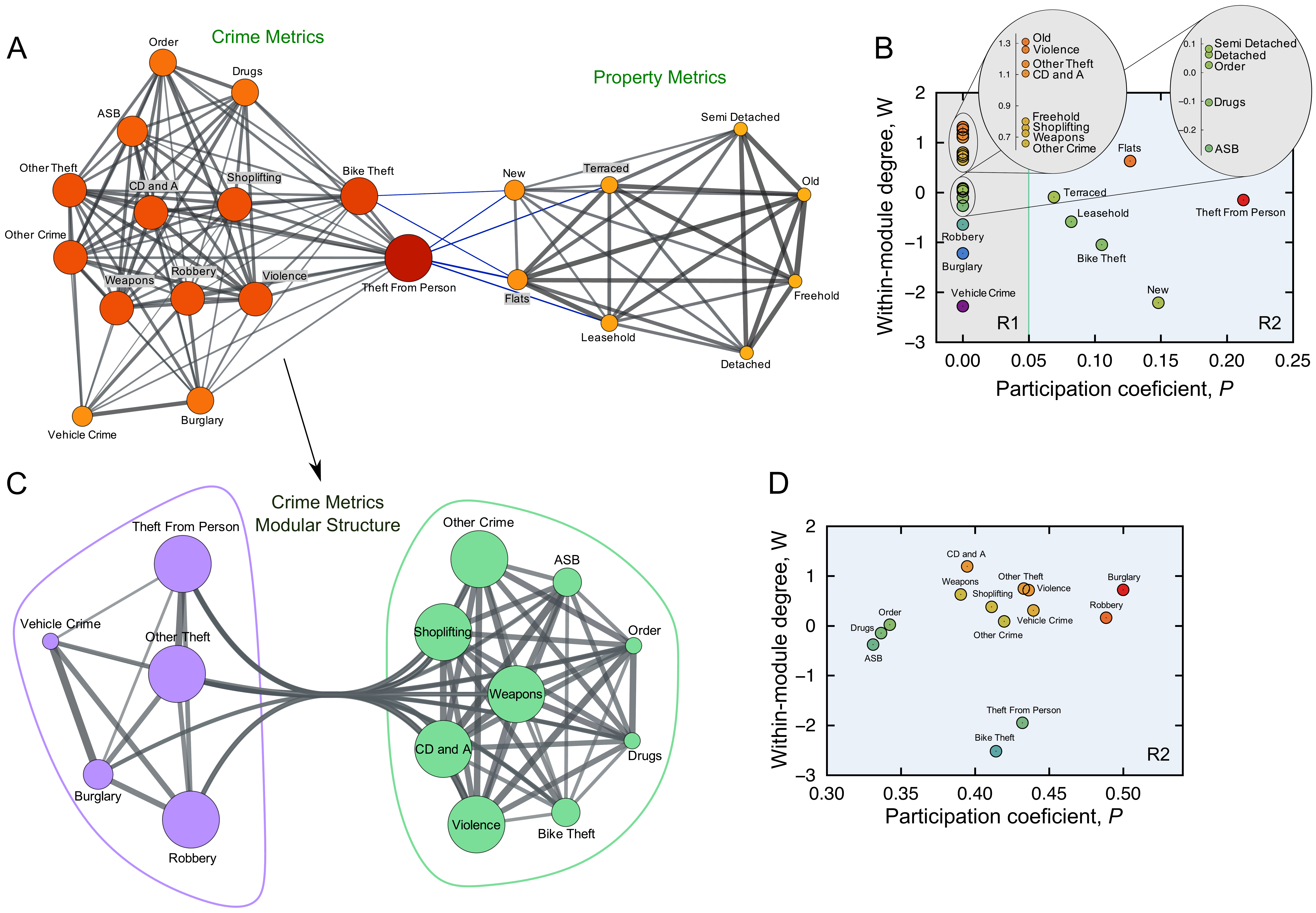}
\end{center}
\caption{
\textbf{Network of DSAMs that are positively correlated.} (a) Complex network representation of the positive connections among crime and property DSAMs. Each node is a crime or property type and the connection between two nodes occurs whenever there is a statistically significant correlation between their DSAMs (based on bootstrapping the Pearson correlation and 99\% confidence). Each connection is weighted by the Pearson correlation coefficient and the thickness of the edges are proportional to the connection weight. Node sizes are proportional to their degrees and the color code also refers to node degree. A modular structure composed of two modules (one with all property metrics and a second with all crime metrics) is identified by maximizing the network modularity (yielding $M=0.47$ for the original network and $\langle M_{\text{rand}}\rangle=0.12\pm0.01$ for a set of randomizations of the original network). Edges highlighted in blue are ones connecting the two modules. (b) Characterization of nodes based on the within-module connectivity ($W$) and participation coefficient ($P$). Each dot in the $W$-$P$ plane corresponds to a crime or property type. All nodes are classified as ultraperipheral ($R1$) or peripheral ($R2$); in particular, the majority of nodes has zero participation coefficient (that is, has only within-module links) and only the six nodes in the $R2$ region have between modules connections. (c) Modular structure of the sub-graph related to the crime metrics. For this case, two modules (colored in purple and green) are found by maximizing the network modularity ($M=0.14$ and $\langle M_{\text{rand}}\rangle=0.06\pm0.01$). (d) Role discrimination of crime nodes by the $W$-$P$ plane. We note that all nodes are in the peripheral region ($R2$). Drugs, order, and anti-social behavior (ASB) crime types are the most peripheral; robbery and burglary have the largest $P$, and criminal damage and arson (CD and A) has the largest $W$.
}
\label{fig:5}
\end{adjustwidth}
\end{figure*}

In order to build these complex networks, we bootstrap the Pearson correlation, $\rho_{ij}$, for every pair of metrics (over one thousand realizations), identifying those that are statistically significant at 99\% confidence level. The significant correlations are shown in~\hyperref[S6_Fig]{S6 Fig}, where we can individually visualize the effect of all crime and property categories on a particular one. Next, we group all pairs of metrics having significant positive correlations to create the weighted complex network of Figure~\ref{fig:5}A. In this representation, the vertices are crime and property categories, the edges indicate the existence of significant positive correlations, and the edge weights are the correlation values. 

We apply the network cartography of Rimerà and Amaral~\cite{guimera2005functional,guimera2005cartography} to extract the network modules and classify nodes according to their within- ($W$, in standard score units) and between-module connectivity (or participation coefficient $P$, a fraction). This approach yields the same two main modules observed in the hierarchical clustering, that is, a crime and a property module. We find the significance of the this modular structure by comparing the network modularity $M$ (the fraction of within-module edges minus the fraction expected by random connections~\cite{guimera2005functional,guimera2005cartography,newman2004finding,newman2004fast}) with the average modularity $\langle M_{\text{rand}}\rangle$ of randomized versions of the original network~\cite{guimera2004modularity}. For these modules, we have $M=0.47$ and $\langle M_{\text{rand}}\rangle = 0.12 \pm 0.01$, showing that the modular structure cannot be explained by chance. Figure~\ref{fig:5}B shows a classification of the crime and property categories based on the $W$-$P$ plane (within-module connectivity vs. between-module connectivity). We note that most metrics have $P=0$, that is, these metrics only have  within-module connections (ultraperipheral nodes $R1$ according to~\cite{guimera2005functional,guimera2005cartography}). Weak positive correlations exist between the crime types: bike theft and theft from the person, and the property categories: flats, leasehold, new, and terraced. Within each module, we find violence and other theft to be the most connected categories in the crime module; while old and freehold are the most connected types in the property module. These crime and property types are expected to have the largest positive impact on their modules, meaning that an increase/decrease in their DSAM values correlates to an increase/decrease in several other types within their modules.

We also ask if these modular structures can be broken into sub-modules. To answer this question, we apply the network cartography to the two sub-graphs composed by the crime and property modules. For the property module, no significant sub-modular structure could be found ($M=0.12$ and $\langle M_{\text{rand}}\rangle = 0.12 \pm 0.05$). For the crime module, the sub-modular structure shown in Figure~\ref{fig:5}C is significant ($M=0.14$ and $\langle M_{\text{rand}}\rangle = 0.06 \pm 0.01$). We note the existence of two modules: one (on the left) is dominated by acquisitive types of crime and consists of theft from the person, other theft, robbery, burglary, and vehicle crime; the other contains all remaining categories. We also find that these sub-modules cannot be broken into statistically significant smaller structures. The role discrimination of crime nodes based on the $W$-$P$ plane is shown in Figure~\ref{fig:5}D, where all nodes are classified as peripheral nodes ($R2$ -- see~\cite{guimera2005functional,guimera2005cartography}), which reflects the entanglement among crime types. In spite of that, we find burglary and robbery to be the most interconnected categories (that is, having the largest $P$); while anti-social behavior (ASB), drugs and order are the most ``local'' categories. Naturally, correlation does not imply causation and our analysis must be viewed as a seminal alternative proposal for investigating the inter-relationships among different crime types. Taking these points into account, our approach suggests that policies focused on reducing burglary and robbery are more likely to ``spread'' over other crime types than those eventually focused on categories such as anti-social behavior (ASB), drugs and order. This result suggests that actions such as ``the zero-tolerance policies'' against minor crimes with lower participation and connectedness are unlikely to have a strong positive impact on reducing more serious crimes when compared with policies focused on more entangled crime types. 

\begin{figure*}[!ht]
\begin{adjustwidth}{-2.25in}{0in}
\begin{center}
\includegraphics[scale=0.16]{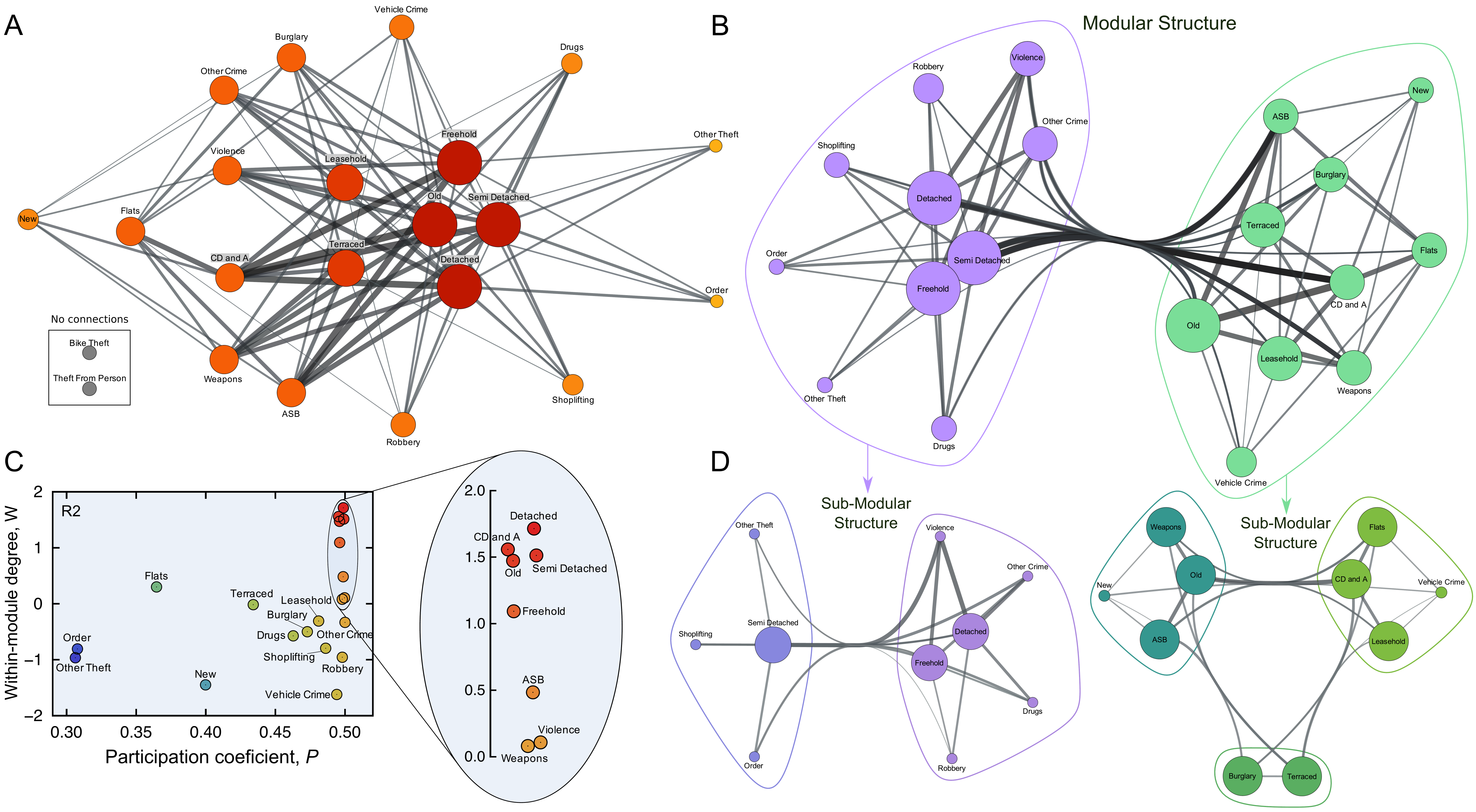}
\end{center}
\caption{
\textbf{Network of DSAMs that are negatively correlated.} (a) Complex network representation of the negative correlations among crime and property DSAMs. Each node is a crime or property type and the connection between two nodes occurs whenever there is a statistically significant anti-correlation between their DSAMs (based on bootstrapping the Pearson correlation and 99\% confidence). Node sizes are proportional to their degrees and the color code also refers to node degree. Each connection is weighted by the absolute value of the Pearson correlation coefficient and the thickness of the edges are proportional to the connection weight. (b) Modular structure of the negatively correlated network. Two modules are identified by maximizing the network modularity ($M=0.13$ and $\langle M_{\text{rand}}\rangle=0.07\pm0.02$) and are colored in purple and green. (c) Role discrimination of nodes by the $W$-$P$ plane (within-module connectivity versus participation coefficient). We note that all nodes are in the peripheral region ($R2$). (d) Modular structure of the sub-graphs related to the two modules of (b). One of the modules can be divided into two sub-modules that has been colored with purple shades ($M=0.15$ and $\langle M_{\text{rand}}\rangle=0.06\pm0.02$) and the other yields three sub-modules that are colored with green shades ($M=0.14$ and $\langle M_{\text{rand}}\rangle=0.08\pm0.02$). These sub-modular structures reveal that some property types have their values more depreciated by specific crime types.
}
\label{fig:6}
\end{adjustwidth}
\end{figure*}

Analogous to the previous case, we investigated the network of negative correlations. In this representation, we connect every crime and property type displaying significant negative (or anti-) correlations and the edge weights are proportional to the absolute value of these correlations. Figure~\ref{fig:6}A shows that this network has a very distinct structure, where crime types are never connected to each other and the same occurs among property types. This means that the increasing/decreasing of DSAM for a particular crime does not correlate to a decreasing/increasing of DSAM for any other crime category. The same holds for property types. Thus, an increase/decrease of DSAMs for crime types is only correlated to a decrease/increase of DSAMs for property categories, illustrating that criminal activities have an important role in the depreciation process of property values. Interestingly, bike theft and theft from the person deviate from this behavior and have no significant negative correlations to any other metric.

We also apply the network cartography to the network of negative correlations, finding that it can be broken into two significant modules ($M=0.13$ and $\langle M_{\text{rand}}\rangle=0.07\pm0.02$ -- Figure~\ref{fig:6}B). One module is composed by detached, freehold, and semi-detached property types as well as seven crime categories (drugs, order, other crime, other theft, robbery, shoplifting, and violence). The other module is formed by flats, leasehold, new, old, and terraced properties surrounded by the remaining seven crime categories. Figure~\ref{fig:6}C shows the role discrimination of nodes based on the $W$-$P$ plane. As in the sub-modular structure of crime metrics (Figures~\ref{fig:5}C~and~\ref{fig:5}D), all nodes in the network of negative correlations are classified as peripheral nodes ($R2$). This result reinforces the interconnectedness of this network, indicating that is very hard to find crime types having a very uneven impact on property values. 

In spite of these conditions and remembering that our analysis must be viewed as a first step toward a better understanding of the inter-relationships among crime and property types, we observe that detached, old, semi detached and freehold property types have the largest values of $P$ and $W$. This result suggests that these properties are the most susceptible to having their values depreciated by  criminal activities. We also note that anti-social behavior (ASB), criminal damage and arson (CD and A), violence, and weapons have the largest values of $P$, suggesting that these crime types exhibit a distinct influence on the property values; criminal damage and arson (CD and A) also has a large value of $W$, indicating that this crime category has both an influence over its module and over the other module. The most ``local'' crime categories are order and other theft (smallest values of $P$), indicating that they have an important impact only on the property values of their module. Similarly, flats and new properties have the smallest $P$ among property types, suggesting that these properties are most affected by crime types belonging to their module. 

We tested for additional structure and found the modules could be broken into the sub-modules shown in Figure~\ref{fig:5}D. The sub-graph composed by the module on the left of Figure~\ref{fig:5}B yields two sub-modules ($M=0.15$ and $\langle M_{\text{rand}}\rangle=0.06\pm0.02$), while the module on the right of Figure~\ref{fig:5}B yields three sub-modules ($M=0.14$ and $\langle M_{\text{rand}}\rangle=0.08\pm0.02$). Each of these sub-modules is composed by one or two property types and from one (the one composed by burglary and terraced) to four crime categories (the one composed by detached, freehold, drugs, other crime, robbery, and violence). It is not easy to explain such groups or to claim that these sub-modular structures are very meaningful since the original network and its modular structure is very entangled (which is quantified by the small values of the modularity $M$). However, the statistical significance of these structures suggests the depreciation process of property values caused by criminal activities is hierarchically organized.

\section*{Conclusion}
This study advances our understanding of the inter-relationship between police reported crime and property transaction values using density scale-adjusted metrics. When the trend attributable to population density is removed using allometric scaling laws, the resulting metrics more effectively compare constituencies. This study reaches a number of important conclusions.

Individual categories of DSAMs may appear to exhibit no trends and be consistent with a normal distribution, however, when looking at single indicators, important and significant correlations will remain unobserved. In the current study, DSAMs were observed to exhibit significant positive and negative correlations with a host of other metrics. 

Correlations between DSAMs from different crime indicators revealed universally positive correlations with every other crime indicator. Similarly, density scale-adjusted metrics for property transaction values were positively correlated with all other property types. These results indicate that at the level of parliamentary constituencies an increase in the DSAM for one type of crime predicts an increase in all other types of crime. It should be noted, that DSAMs will account for general rises and falls in crime across all scales. Thus, a decrease in absolute numbers does not mean the scale-adjusted metric will decrease.

With the exceptions of bike theft and theft from the person, crime and property DSAMs are negatively correlated. This means that as a general rule, an increase in DSAM of crime is associated with a decrease in the value of property transactions. Two crime categories exhibit a particularly strong effect: anti-social behavior (ASB) and criminal damage and arson (CD and A). This indicates that in our data twelve out of fourteen crime types show no evidence of crime targeting affluence. Our network approach also revealed that crime and property DSAMs form hierarchically-organized structures with statistically significant modular and sub-modular structures. These structures represent the crime and/or property categories that are more likely to affect each other. Consequently, such groups may help policy-makers to design more effective actions for reducing crime incidence, with the advantage of having an approach that works over the full range of human environments. 

\clearpage

\section*{Authors Contributions}
H.V.R. and Q.H. conceived and designed the study, carried out the statistical analysis and drafted the manuscript; D.L. conceived the study and helped draft the manuscript. All authors gave final approval for publication.

\section*{Competing interests}
Dan Lewis is the Chief Executive of the Economic Policy Centre (EPC) and Director of UKCrimeStats. Views discussed in the manuscript do not represent the views or positions of UKCrimeStats and EPC. There are no patents, products in development or marketed products to declare. This does not alter our adherence to  PLOS ONE policies on sharing data and materials. 

\section*{Funding}
HVR thanks the financial support of CNPq (grant 440650/2014-3). The funders had no role in study design, data collection and analysis, decision to publish, or preparation of the manuscript.

\section*{Data Availability}
All necessary data to fully reproduce the results of this article are public, freely available,  and has been provided as a spreadsheet for publication with the paper.


\bibliography{density_sam}
\clearpage
\begin{adjustwidth}{-2.25in}{0in}
\section*{Supporting Information}


\subsection*{S1 Dataset}
\label{S1_Dataset}
{\bf Data employed in this study.} Snapshot of police reported crime captured 10/6/2015 and property transaction values captured 17/7/2015 for the 12 months of 2014.

\subsection*{S1 File}
\label{S1_file}
{\bf Visualization of the DSAMs for all Parliamentary Constituencies and for each crime and property type.} %
\clearpage

\subsection*{S1 Fig}
\textbf{Normality tests for DSAMs.} The panels show the cumulative distributions of the normalized DSAMs (that is, after subtracting its mean and dividing by its standard deviation) for each crime and property type. For DSAMs in which the double power-law model is a better fit, we show the distributions of the normalized DSAM defined with the single power-law model (blue lines) and with double power-law model (red lines). The insets in each plot show the $p$-values of the Kolmogorov-Smirnov normality test, where the horizontal dashed lines indicate the 0.01 confidence level threshold. We note that the normality of the DSAMs is achieved for criminal damage and arson (CD and A), detached, freehold and semi detached only when defining the DSAM with the double power-law model. We further observe that double power-law usually produces higher $p$-values.
\label{S1_Fig}

\begin{center}
\includegraphics[scale=0.21]{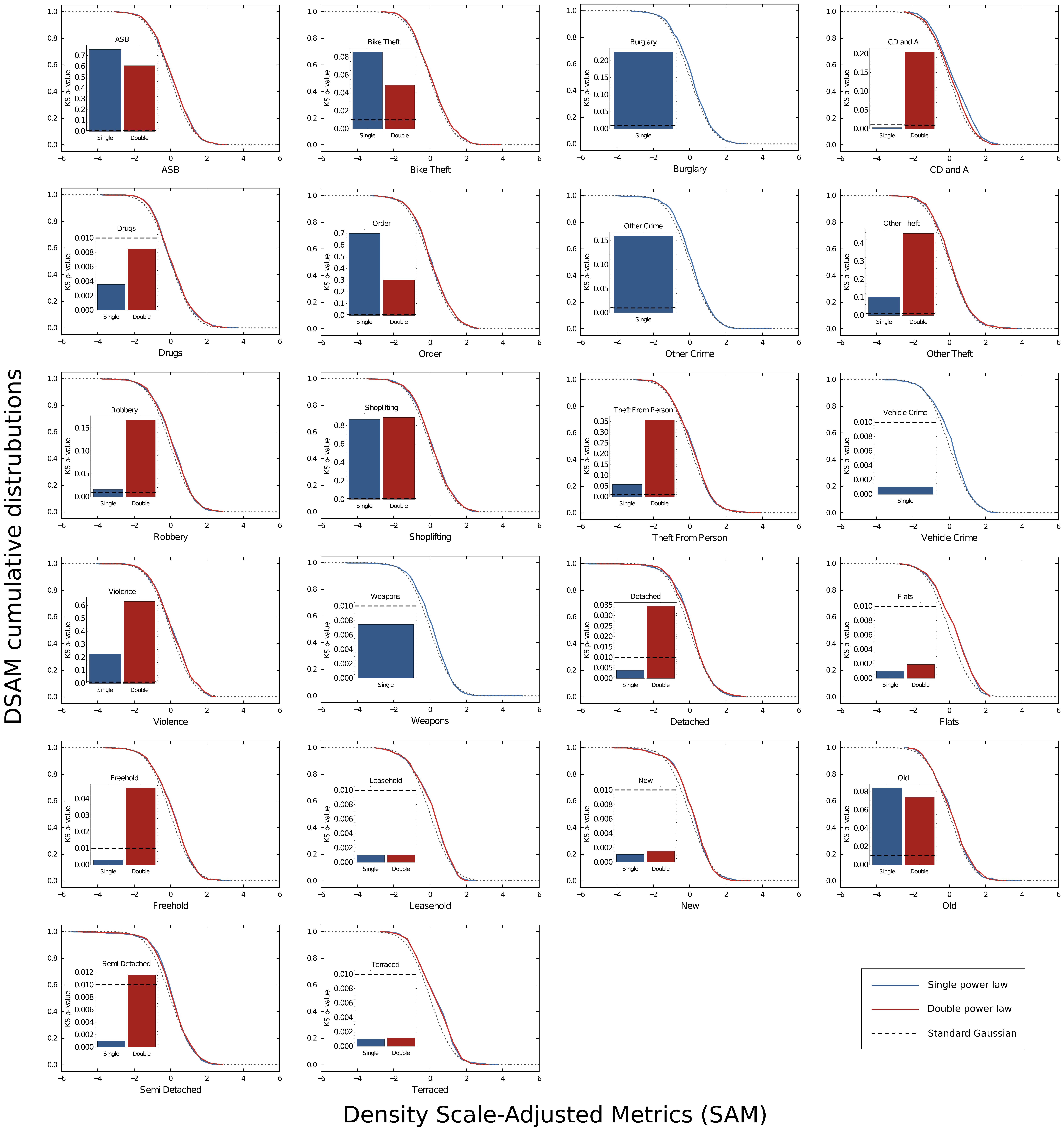}
\end{center}
\clearpage

\subsection*{S2 Fig}
\textbf{Relationship between DSAM standard deviation and log of population density.} Each panel shows the standard deviation of the DSAM evaluated in 15 equally-spaced windows over population density on a logarithmic scale. The error bars are 99\% bootstrap confidence intervals and the horizontal lines are the average values of standard deviations for each indicator. The plots with blue markers are the ones for which no significant increasing trend is observed between standard deviation and population density via linear regressions, whereas the plots with red markers show the indicators for which this relationship has a significant linear increasing trend. Despite the statistical significance of some linear coefficients (9 out of 22), we observe that the majority of the relationships do not show large deviations from the horizontal plateau defined by the mean of the standard deviation.
\label{S2_Fig}

\begin{center}
\includegraphics[scale=0.3]{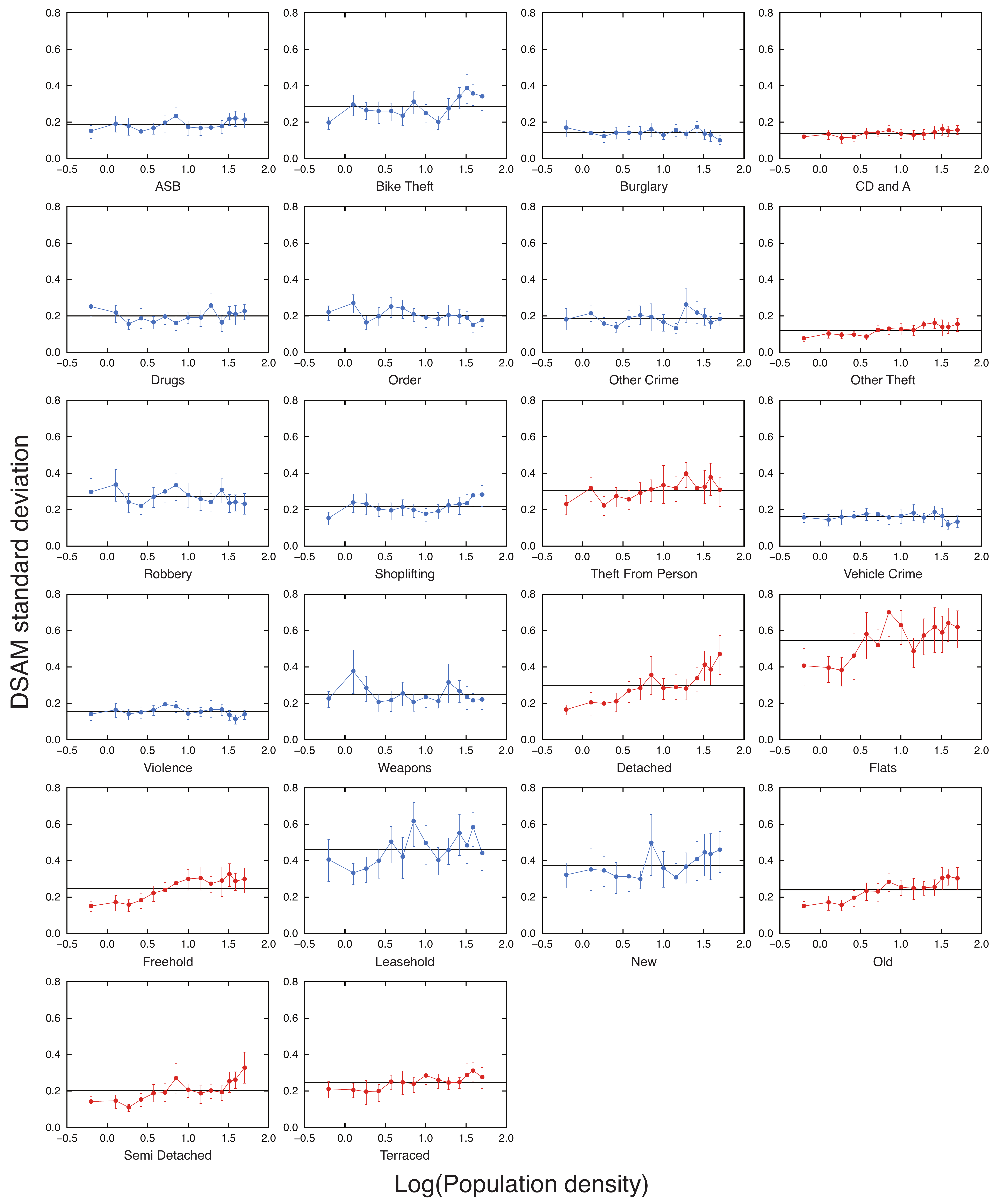}
\end{center}
\clearpage

\subsection*{S3 Fig}
\textbf{Quantifying the relationship between DSAM standard deviation and log of population density.} Linear coefficients of the linear regressions between DSAM standard deviation and log of population density that are statistically significant. Despite the significance of the increasing trends for 9 out of 22 metrics, we note that the growth rates are very small (up to 0.005 units of DSAM per log[p/ha]). 
\label{S3_Fig}

\begin{center}
\includegraphics[scale=0.3]{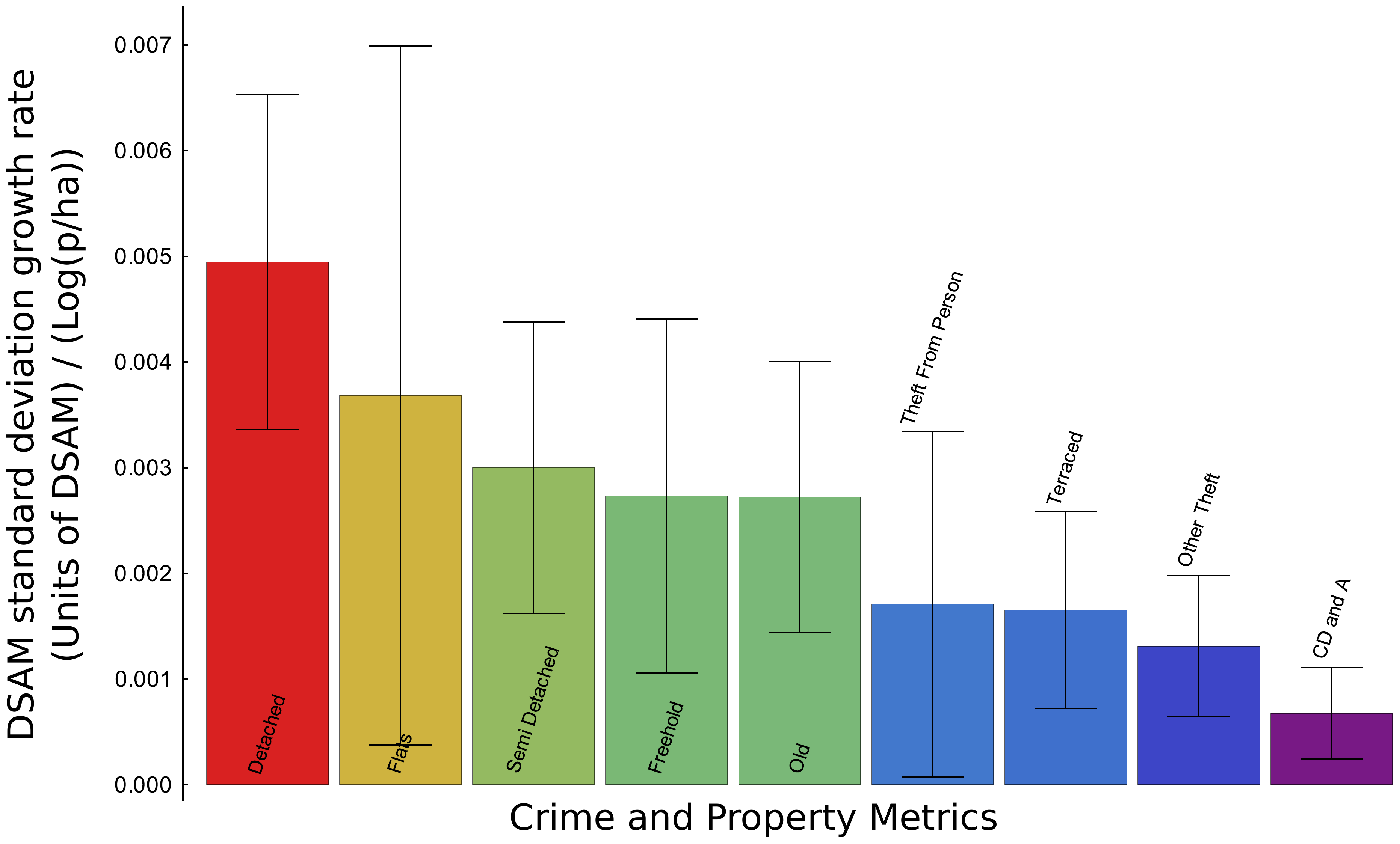}
\end{center}
\clearpage

\subsection*{S4 Fig}
\textbf{Quantifying the degree of nonlinearity in the DSAMs cross-corrections with MIC.} The bar plot show the average value of the difference between the maximal information coefficient (MIC)~\cite{Reshef} and square of the Pearson linear correlation over over all unique pairs of DSAM types (MIC-Pearson$^2$), for the DSAMs values after shuffling their values among constituencies, and for uniform random numbers (sample size is equal to the DSAM case). Error bars are one standard deviation of the quantity (MIC-Pearson$^2$). We note that the average value for original DSAMs is small and fully explained by chance. Thus, we can assume that correlations among the DSAMs are linear. 
\label{S4_Fig}

\begin{center}
\includegraphics[scale=0.3]{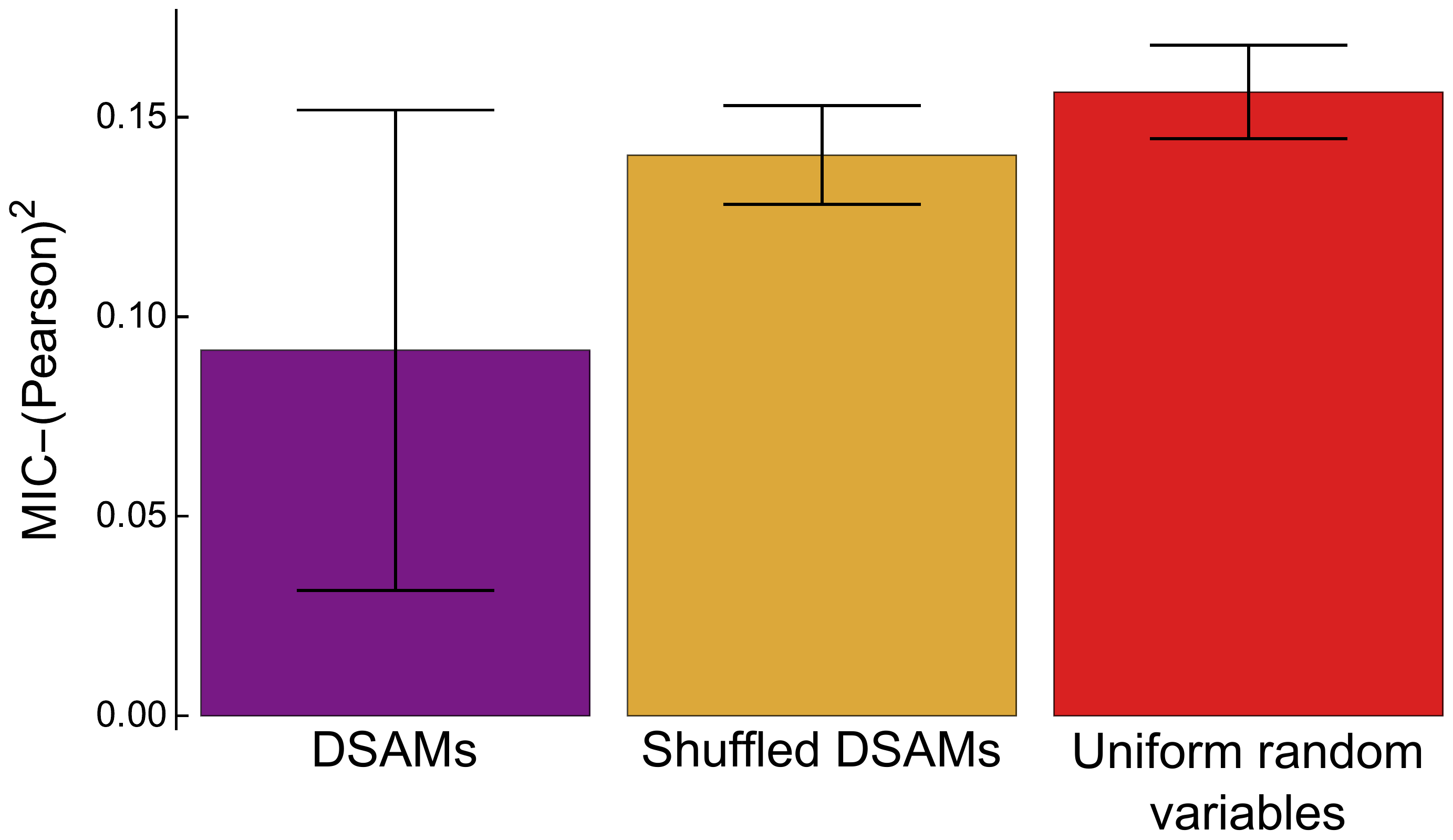}
\end{center}

\clearpage
\subsection*{S5 Fig}
\textbf{Quantifying the degree of nonlinearity in the DSAMs cross-corrections by comparing the linear model with quadratic and cubic models.} (A) Each dot corresponds to the $p$-value of the two-sample bootstrap test (at 95\% confidence)for the equality of mean values of the Bayesian information criterion (BIC) obtained by adjusting the DSAMs pairwise relationships with a linear and a quadratic model. We have bootstrapped the BIC values over 100 realizations and considered only the pairwise relationships in which the BIC for quadratic model is smaller than the one obtained for the linear model. Among the 231 possible relationships, the BIC of the quadratic model is smaller than BIC of the linear in 90 cases. However, the difference between the BIC values is significant only in 19 cases (indicated in the plot), that is, only in $\approx$8\% of all relationships. (B) The same analysis comparing the linear model with the cubic model. In this case, the BIC of the cubic model is smaller than BIC of the linear in 71 cases, but in only 23 there is significant difference, only in $\approx$10\% of all relationships. 
\begin{center}
\includegraphics[scale=0.6]{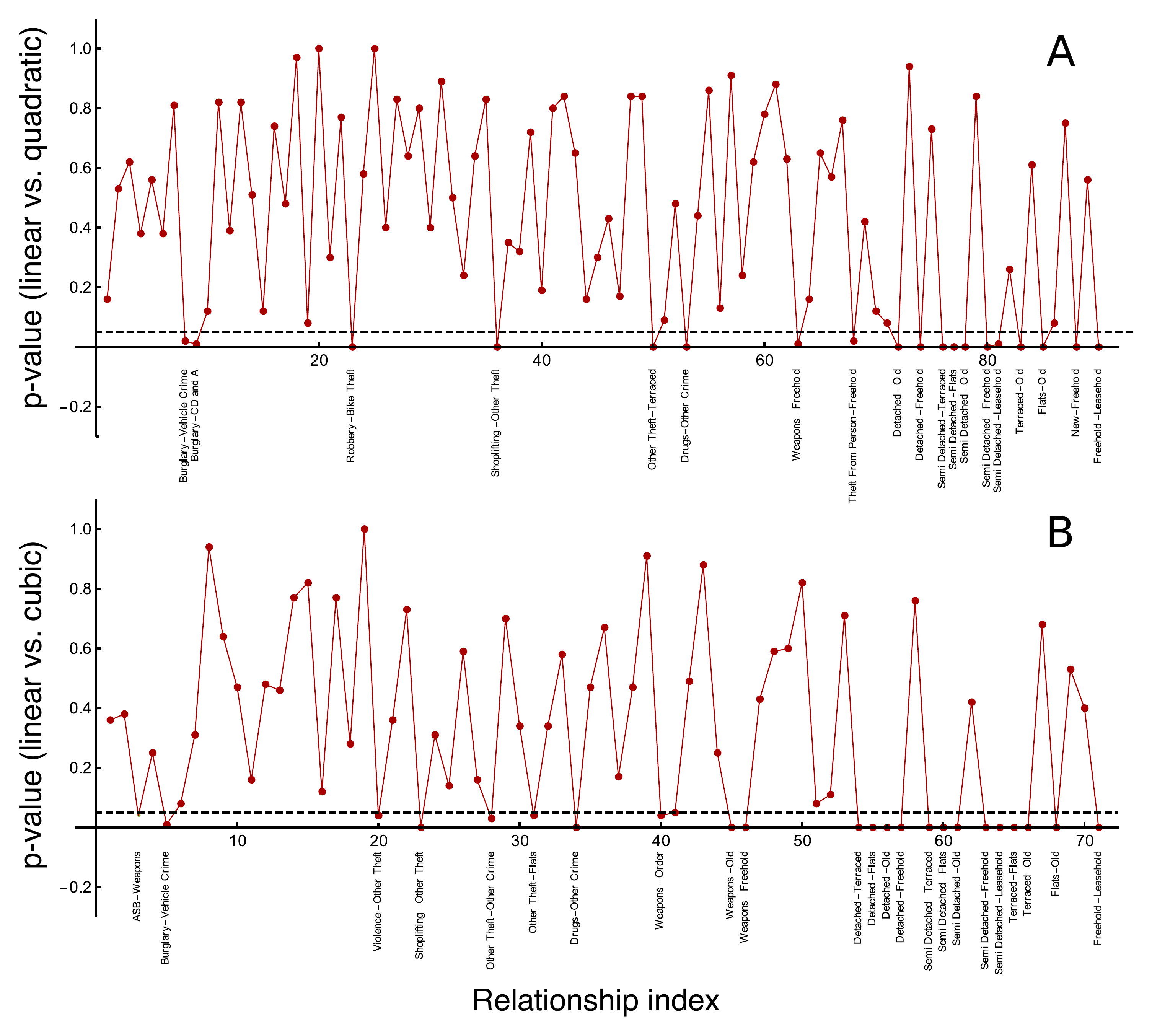}
\end{center}

\clearpage
\subsection*{S6 Fig}
\textbf{Individual visualization of the node connections in the positive and negative correlated networks.} Each gray node (circle) represents a crime or a property type, and the connections indicate the metrics that are negatively (blue) and positively (red) correlated with the particular crime or property type. The edge thickness is proportional to the absolute value of Pearson correlation $\rho_{ij}$ between the metrics, and the edge label shows the value of $\rho_{ij}$. The edges are arranged so that negative correlations are above the gray nodes, and positive ones are below; also, the closer the edge is to the vertical position, the more intense is the correlation/anti-correlation between the metrics. 
\label{S6_Fig}

\begin{center}
\includegraphics[scale=0.15]{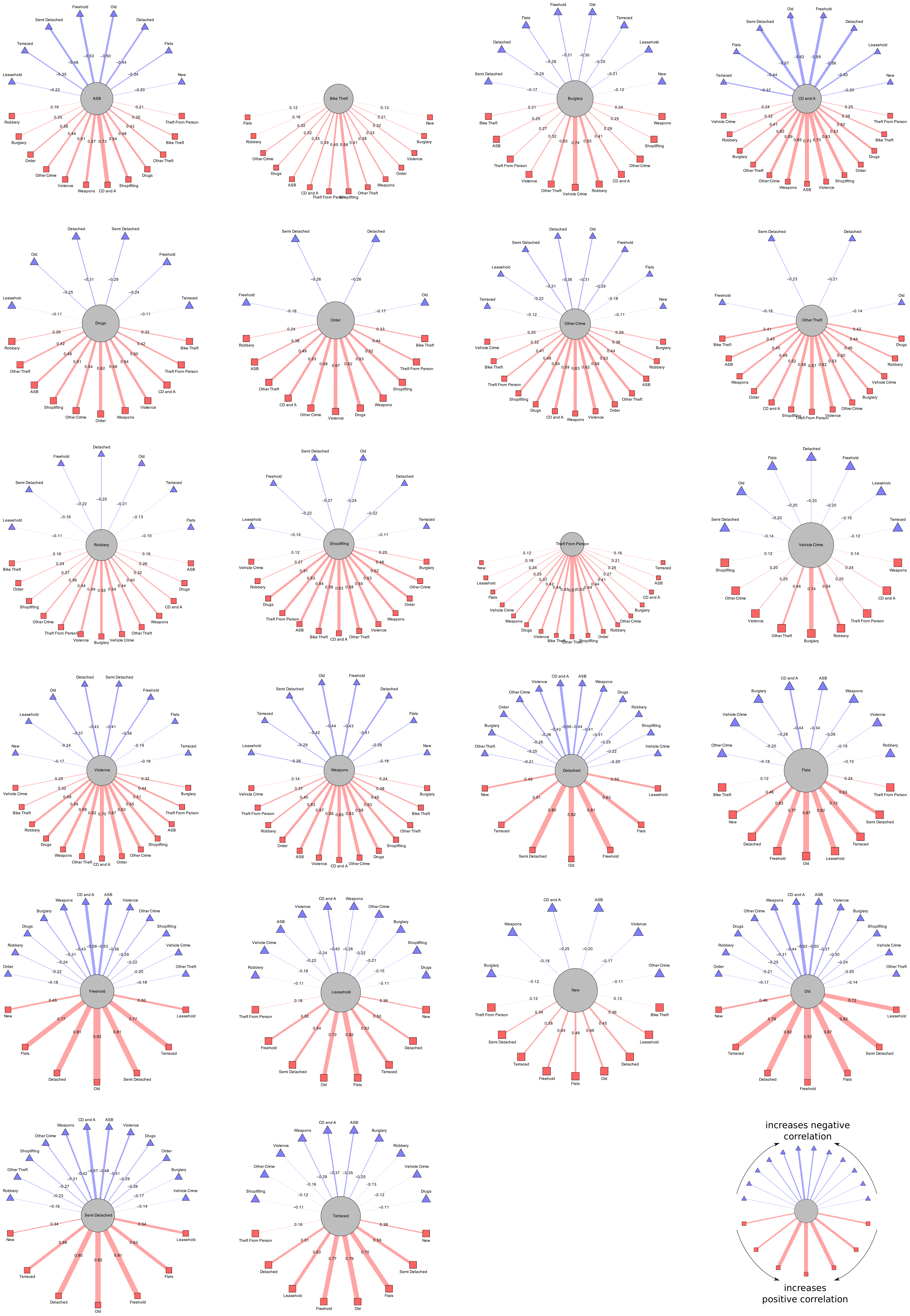}
\end{center}

\end{adjustwidth}

\end{document}